\documentclass[10pt]{article}
\usepackage{multicol}
\usepackage{graphicx}
\usepackage{amsmath}
\usepackage[a4paper]{geometry}
\usepackage{hyperref}
\usepackage{rotating}

\begin{document}

\begin{center}
{\Large \bf Effects of coalescence and isospin symmetry on the
freezeout of light nuclei and their anti-particles}

\vskip1.0cm

M.~Waqas$^{1,}${\footnote{Corresponding author. Email (M.Waqas):
waqas\_phy313@yahoo.com; waqas\_phy313@ucas.ac.cn}},G. X.
Peng$^{1,2,3}$ {\footnote{Corresponding author. Email (G. X.
Peng): gxpeng@ucas.ac.cn}}, Fu-Hu Liu$^{4,5}$ {(fuhuliu@163.com;
fuhuliu@sxu.edu.cn)}, Z.~Wazir$^{6}${(zwazir@gudgk.edu.pk)}
\\

{\small\it  $^1$ School of Nuclear Science and Technology, University of Chinese Academy of Sciences,
Beijing 100049, China,

$^2$ Theoretical Physics Center for Science Facilities, Institute of High Energy Physics, Beijing 100049, China,

$^3$ Synergetic Innovation Center for Quantum Effects \&
Applications, Hunan Normal University, Changsha 410081, China,
$^4$Institute of Theoretical Physics \& State Key Laboratory of
Quantum Optics and Quantum Optics Devices, Shanxi University,
Taiyuan, 030006, China,

$^5$Collaborative Innovation Center of Extreme Optics, Shanxi
University, Taiyuan, 030006, China,

$^6$Department of physics, Ghazi University, Dera Ghazi Khan,
Pakistan.}

\end{center}

\vskip1.0cm

{\bf Abstract:} The transverse momentum spectra of light nuclei
(deuteron, triton and helion)
 produced in various centrality intervals in Gold-Gold (Au-Au), Lead-Lead (Pb-Pb) and proton-Lead (p-Pb) collisions, as well as in inelastic
 (INEL) proton-proton (pp) collisions are analyzed by the blast wave model with Boltzmann Gibbs statistics. The model
 results are nearly in agreement with the experimental data measured by STAR and ALICE
 Collaborations in special transverse momentum ranges. We extracted the bulk properties in terms of
 kinetic freezeout temperature, transverse flow velocity and freezeout volume. It is observed that deuteron and anti-deuteron freezeout later
than triton and  helion as well as their anti-particles due to its
smaller mass, while helion and triton, and anti-helion and
anti-triton freezeout at the same time due to isospin symmetry at
higher energies. It is also observed that light nuclei freezeout
earlier than their anti-nuclei due to the large coalescence of
nucleons for light nuclei compared to their anti-nuclei. The
kinetic freezeout temperature, transverse flow velocity and
kinetic freezeout volume decrease from central to peripheral
collisions. Furthermore, the transverse flow velocity depends on
mass of the particle which decreases with increasing the mass of
the particle.
\\



\vskip1.0cm

\begin{multicols}{2}

{\section{Introduction}} A new form of matter, the so called
quark-gluon plasma (QGP) is produced under the high temperatures
and energy densities in relativistic heavy ion collisions. This
matter is formed in the early stage of collisions that survives
for a very short period of time ($\sim$ 7-10 fm/c), after which
the QGP gets transformed rapidly to a system of hadron gas. Due to
multi-partonic interactions throughout the evolution time in the
collisions, the information about the initial condition of the
system get mostly lost. The final state behavior of such colliding
system can be attained from the measurement of the number as well
as the identity of the produced particles along with their energy
and momentum spectra. The final state information is very useful
to understand the particle production mechanisms and the nature of
the matter created in these high energy collisions.

Temperature is one of the most crucial factor in sub-atomic
physics. There are different types of temperatures present in
literature [1--5]. Chemical freezeout temperature ($T_{ch}$),
which describes the excitation degree of interacting system at the
stage of chemical freezeout. At chemical freezeout the chemical
components (relative fraction) of the particles are invariant. The
excitation degree of the interacting system at the stage of
thermal or kinetic freezeout is described by the kinetic freezeout
temperature ($T_0$). At kinetic freezeout, the transverse momentum
spectra of the particles are no longer changed. Another type of
temperature is the effective ($T_{eff}$). It is not a real
temperature but is related to the particle mass and can be
extracted from the transverse momentum spectra by using various
distribution laws such as standard (Boltzmann, Bose-Einstein and
Fermi-Dirac), Tsallis, and so forth.

The chemical freezeout temperature is usually obtained from the
particle ratio [6--8]. Due to chemical equilibrium being meanwhile
or earlier than the kinetic equilibrium, the chemical freezeout
temperature is equal or higher than the kinetic freezeout
temperature. The effective temperature, due to mass and flow
velocity are also larger than the kinetic freezeout temperature
[9, 10]. Due to more violent interactions in central collisions,
both the chemical freezeout temperature and effective temperature
are larger in central collisions than in the peripheral
collisions. However, the situation for kinetic freezeout
temperature is not clear. several literatures claim larger $T_0$
in central collisions [11-16] which decrease towards periphery,
while several claims larger $T_0$ in peripheral collisions
[17--20] which decrease towards the central collisions. In
addition, volume is also very important parameter in high energy
collisions. The volume occupied by the ejectiles when the
correlative interactions become negligable and the only force they
experience is Coulombic force, is said to be kinetic freezeout
volume ($V$). Most of the literatures agreed to the larger $V$ as
well as the transverse flow velocity ($\beta_T$) in central
collisions which decrease towards periphery.

Freezeout scenario is very important in high energy collisions.
Different freezeout scenarios are discussed in literature at
different stages of the freezeout, but we will focus on kinetic
freezeout scenarios in the present work. There are several kinetic
freezeout scenarios in literature [15, 16, 21--24] which include
single, double, triple and multiple freezeout scenario. In the
study of production of light nuclei, it is expected that the
freezeout of the particles may also be dependent on the nucleon
coalescence and isospin symmetry at higher energies.

The transverse momentum spectra ($p_T$) of the particles are very
important observables because they give very crucial information
about the equilibrium dynamics and the anisotropy of the produced
system in heavy ion collisions [24]. In the present work, we will
analyze the $p_T$ spectra of deuteron ($d$), anti-deuteron ($\bar
d$), triton ($t$), anti-triton ($\bar t$), helion ($^3He$) and
anti-helion ($\bar {^3He}$) in Gold-Gold (Au-Au), Lead-Lead
(Pb-Pb), Proton-Lead (p-Pb) and proton-proton (pp) collisions.
\\

{\section{The method and formalism}} A few methods can be used for
the extraction of $T_0$ and $\beta_T$, including but not limited
to, (1) the blast-wave model with Boltzmann-Gibbs statistics
(BGBW) [26--28], (2) the blast-wave model with Tsallis statistics
[29], (3) an alternative method using the Boltzmann distribution
[22, 28, 30--35], and (4) the alternative method using the Tsallis
distribution [36, 37]. It is noteworthy that $T_0$ is the
intercept in the linear relation $T$-$m_0$ in alternative method,
where $m_0$ is the rest mass; and $\beta_T$ is the slope in the
linear relation $<p_T>$-$\bar m$, where $<p_T>$ is the mean
transverse momentum and $\bar m$ is the mean moving mass (i.e.,
the mean energy).

Ref. [38] confirms that the above methods are harmonious. Among
these methods, the first method is the most direct with fewer
parameters, though it has been revised in various ways and applied
to other quantities [39--43]. We have used the first method, i.e.,
BGBW in the present work. Due to their coherence, other models
will not be used [38].

BGBW is a phenomenological model which is used for the spectra of
hadrons based on flowing local thermal sources with global
variables of temperature, volume and transverse flow velocity.
According to [26--28], the $p_T$ distribution of BGBW can be
written as

\begin{align}
f(p_T)=&\frac{1}{N}\frac{dN}{dp_T} = \frac{1}{N}\frac{gV}{(2\pi)^2} p_T m_T \int_0^R rdr \nonumber\\
& \times I_0 \bigg[\frac{p_T \sinh(\rho)}{T} \bigg] K_1
\bigg[\frac{m_T \cosh(\rho)}{T} \bigg],
\end{align}
where $N$ is the number of particles, $g$ represents the
degeneracy factor of the particle (which is different for
different particles, based on $g_n$=2$S_n$+1, while $S_n$ is the
spin of the particle), $V$ is the freezeout volume, $m_T$ is the
transverse mass ($m_T=\sqrt{p_T^2+m_0^2}$), $I_0$ and $K_1$ are
the modified Bessel functions, $\rho= \tanh^{-1} [\beta(r)]$,
$\beta(r)= \beta_S(r/R)^{n_0}$ is the transverse radial flow of
the thermal source at radius $0 \leq \ r \leq \ $$R$ with surface
velocity $\beta_S$ and $n_0$=1 [29]. In general,
$\beta_T=(2/R^2)\int_0^R r\beta(r)dr=2\beta_S/(n_0+2)=2\beta_S/3$.

Eqs. (1) can be used for the fitting of $p_T$ spectra to obtain
the parameters $T_0$, $V$ and $\beta_T$. It should be noted that
Eqs. (1) can be only valid in a narrow $p_T$ range i.e: they
describe only soft excitation process. However we have to consider
the hard scattering process for the spectra in a wide $p_T$ range.
In general, the contribution of hard process can be parameterized
to an inverse power law [44--46], i.e., Hagedorn function [47,48]

\begin{align}
f_0(p_T)=\frac{1}{N}\frac{dN}{dp_T}= Ap_T \bigg( 1+\frac{p_T}{
p_0} \bigg)^{-n},
\end{align}
which is resulted from the calculus of quantum chromodynamics
(QCD) [44, 45, 46], where $p_0$ and $n$ are the free parameters
and $A$ is the normalization constant that depends on $p_0$ and
$n$.

In order to describe a wide $p_T$ range, the superposition of both
the soft and hard process can be used, which is
\begin{align}
f_0(p_T)=kf_S(p_T) +(1-k)f_H(p_T),
\end{align}
where $k$ shows the contribution fraction of the first component
(soft excitation), while $(1-k)$ represents the contribution
fraction of the second component (hard scattering) in Eq. 3, and
according to Hagedorn model [48] the usual step function can be
also be used for the superposition of soft and hard components.

According to Hagedorn thermal model [48], the two-component BGBW
distribution function can also be structured by using the usual
step function,
\begin{align}
f_0(p_T)=\frac{1}{N}\frac{dN}{dp_T}=A_1\theta(p_1-p_T)f(p_T) \nonumber\\
+ A_2 \theta(p_T-p_1) f(p_T),
\end{align}
where $A_1$ and $A_2$ are the constants which give the two
components to be equal to each other at $p_T$ = $p_1$.
\\

{\section{Results and discussion}} The transverse momentum ($p_T$)
spectra, $(1/2\pi p_T)d^2N/dp_Tdy$, of $d$, $\bar d$ and $t$
produced in Au-Au collisions at 54.4 GeV are analyzed by BGBW
statistics in different centrality classes are demonstrated in
fig. 1. The symbols represent the experimental data of the STAR
Collaboration measured in the mid-rapidity range $|y|<0.5$ and the
solid curve are the results of our fitting by using Eq. (1). The
well approximate description of the model result to the
experimental data of the STAR Collaboration [49] in the special
$p_T$ range can be seen. The event centralities and the values of
free parameters are listed in Table 1. The free parameters include
kinetic freezeout temperature ($T_0$), transverse flow velocity
($\beta_T$), kinetic freezeout volume ($V$), normalization
constant ($N_0$), $\chi^2$ and the degree of freedom (dof). Each
panel is followed by the results of its corresponding ratio of the
data/fit. Fig. 2 demonstrates the $p_T$ spectra, $(1/N_{ev})
d^2N/dp_Tdy$ of $d$, $\bar d$, $^3He$ and $\bar {^3He}$ in various
centrality classes in Pb-Pb collisions at 5.02 TeV. The symbols
stands for the experimental data of the ALICE Collaboration [50]
by the Large Hadron Collider (LHC) and the solid curve represent
our fitting results by using the BGBW statistics. In fig. 2 some
spectra are scaled; such as the spectra of $d$ and $\bar d$ in
centrality intervals 5--10\%, 10--20\%, 20--30\%, 30--40\%,
40--50\%, 50--60\%, 60--70\%, 70--80\% and 90--90\% are multiplied
by 1/2, 1/4, 1/8, 1/16, 1/30, 1/40, 1/40, 1/60 and 1/60,
respectively.
\begin{figure*}[htb!]
\begin{center}
\hskip-0.153cm
\includegraphics[width=15cm]{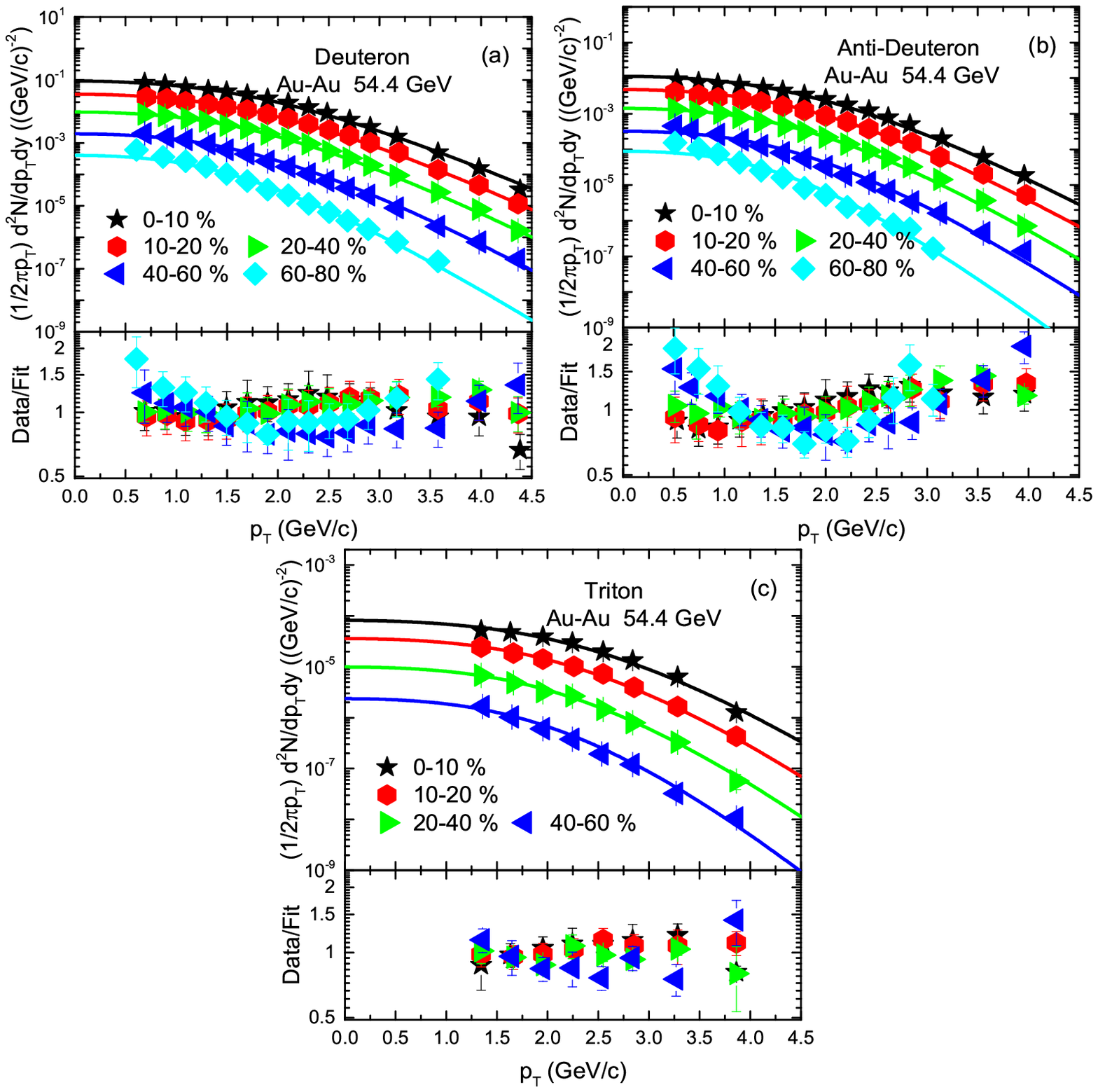}
\end{center}
Fig. 1. The transverse momentum ($p_T$) spectra of (a) $d$ (b)
$\bar d$ and (c) $t$ produced in $|y|<0.5$ in different centrality
classes in Au-Au collisions at 54.4 GeV. The symbols are the
experimental data of the STAR Collaboration [49] and the curves
are our fitting by using BGBW statistics. Each panel is followed
by the ratio of the data/fit.
\end{figure*}

Fig. 3 is similar to fig. 2, but it shows the $p_T$ spectra of
$d$, $\bar d$ and ($^3He$+$\bar {^3He}$)/2 produced in different
centrality intervals in p-Pb collisions at 5.02 TeV. The symbols
show the experimental data measured by the ALICE Collaboration in
the rapidity region $-1 \leq \ y \leq \ $1 and -1$<$$y$$<$0
respectively, and the curves are the results of our fitting by
using Eq. (1). The well approximate description of the model
results to the experimental data of the ALICE Collaboration [51,
52] in the special $p_T$ range can be seen.

In Fig. 4 the $p_T$ spectra of $d$, $\bar d$, $t$, $\bar t$,
$^3He$ and $\bar {^3He}$ in inelastic (INEL) pp collisions at 7
TeV are presented. The symbols represent the experimental data of
the ALICE Collaboration by the LHC in the rapidity interval of
$|y|<0.5$ and the results of our fitting is represented by curve.
The spectra of $\bar d$, $t$, $\bar t$, $^3He$ and $\bar {^3He}$
are multiplied by 1/2.5, 800, 400, 100 and 50, respectively. One
can see the well approximate description of the model results to
the experimental data of the ALICE Collaboration [53] in the
special $p_T$ range.

Fig. 5 shows the variation trend of parameters with centrality
(mass). Panels (a), (b), (c) and (d) show the results for Au-Au
collisions at 54.4 GeV, Pb-Pb collisions at 5.02 TeV, p-Pb
collisions at 5.02 TeV and pp collisions at 7 TeV respectively.
Panels (a), (b) and (c) show the dependence of $T_0$ on
centrality, and panel (d) shows the dependence of $T_0$ on $m_0$.
The types of particles are represented by different symbols. In
Fig. 5, panels (a), (b) and (c), one can see that $d$, $\bar d$,
$t$, $\bar t$, $^3He$ and $\bar {^3He}$ in all collisions (Au-Au,
Pb-Pb and p-Pb ) results in larger $T_0$ in central collisions
which decrease towards periphery. The reason behind this is, in
central collisions, large number of participants involve in
interaction and the collisions are more violent that results in
higher degree of excitation of the system and the kinetic
freezeout temperature is high. However, the collisions become less
violent as the centrality decreases and less number of
participants involve in the interactions which results in
comparatively low kinetic freezeout temperature. This is in
agreement with [11--16], but in disagreement with [17--20].

\begin{figure*}[htb!]
\begin{center}
\hskip-0.153cm
\includegraphics[width=15cm]{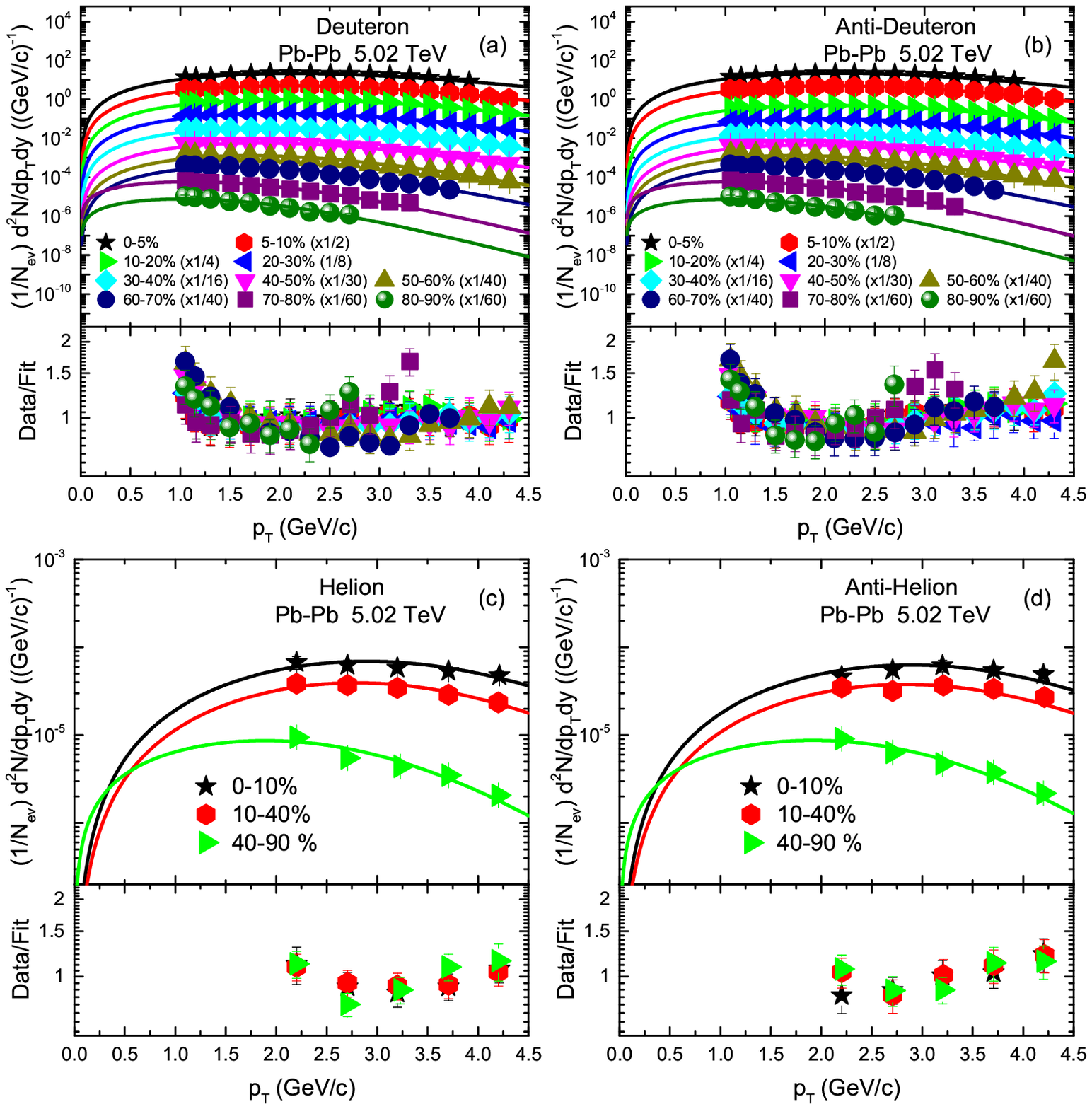}
\end{center}
Fig. 2. Transverse momentum spectra of $d$, $\bar d$, $^3He$ and
$\bar {^3He}$ in $|y|<0.5$ produced in different centrality
intervals in Pb-Pb collisions at 5.02 TeV. The symbols represent
the experimental data measured by the ALICE Collaboration [50],
while the curves are our fitted results by using BGBW statistics,
Eqs. (1). Each panel is followed by the ratio of the data/fit.
\end{figure*}

\begin{figure*}[htb!]
\begin{center}
\hskip-0.153cm
\includegraphics[width=15cm]{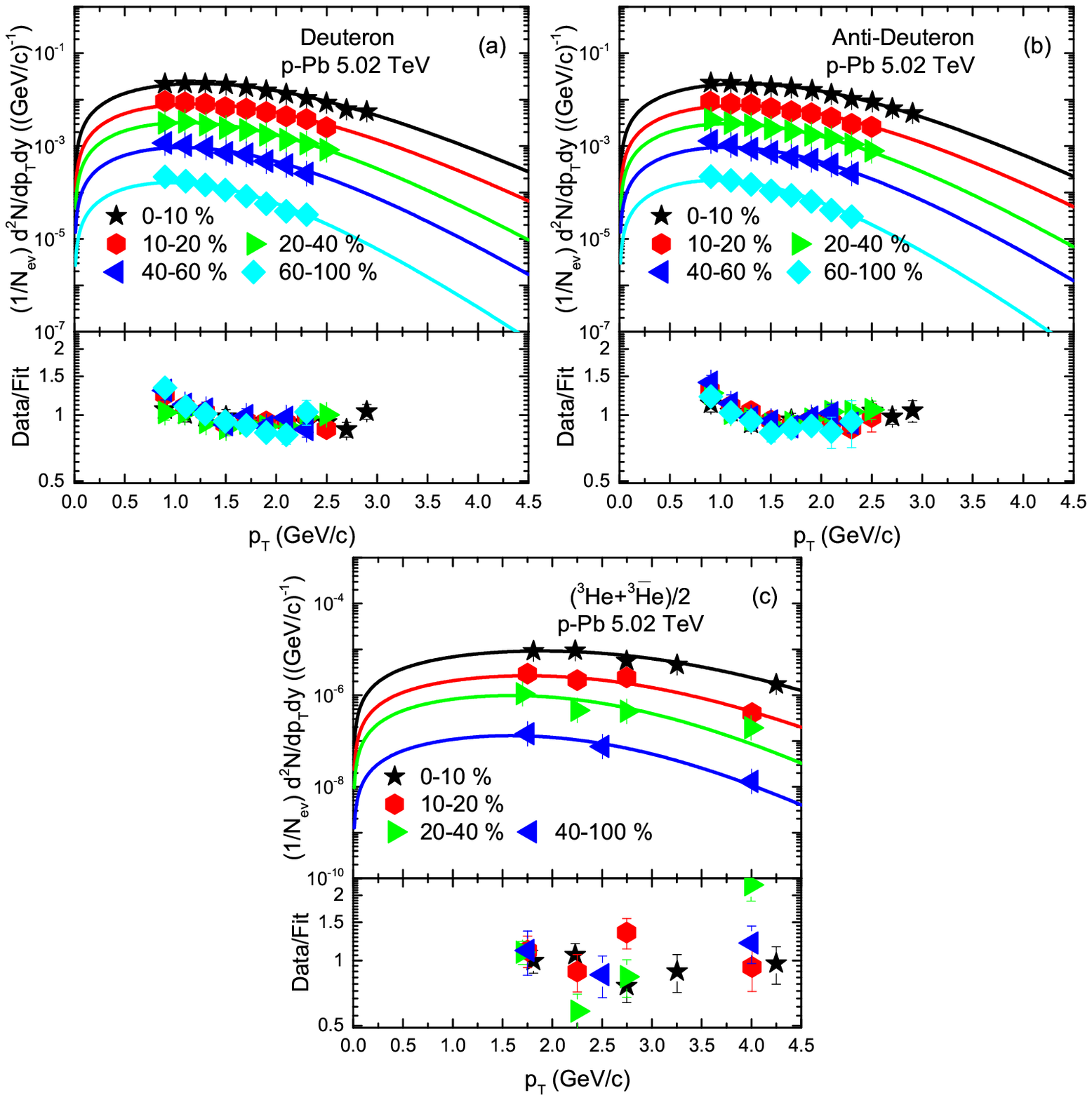}
\end{center}
Fig. 3. Transverse momentum spectra of (a) $d$, (b) $\bar d$ and
(c) ($^3He$+$\bar {^3He}$)/2 produced in various centrality bins
in p-Pb collisions at 5.02 TeV. The symbols represent the
experimental data measured by ALICE Collaboration [51, 52], while
the curves are our fitted results by using BGBW statistics, Eq.
(1). Each panel is followed by the ratio of the data/fit.
\end{figure*}

\begin{figure*}[htb!]
\begin{center}
\hskip-0.153cm
\includegraphics[width=15cm]{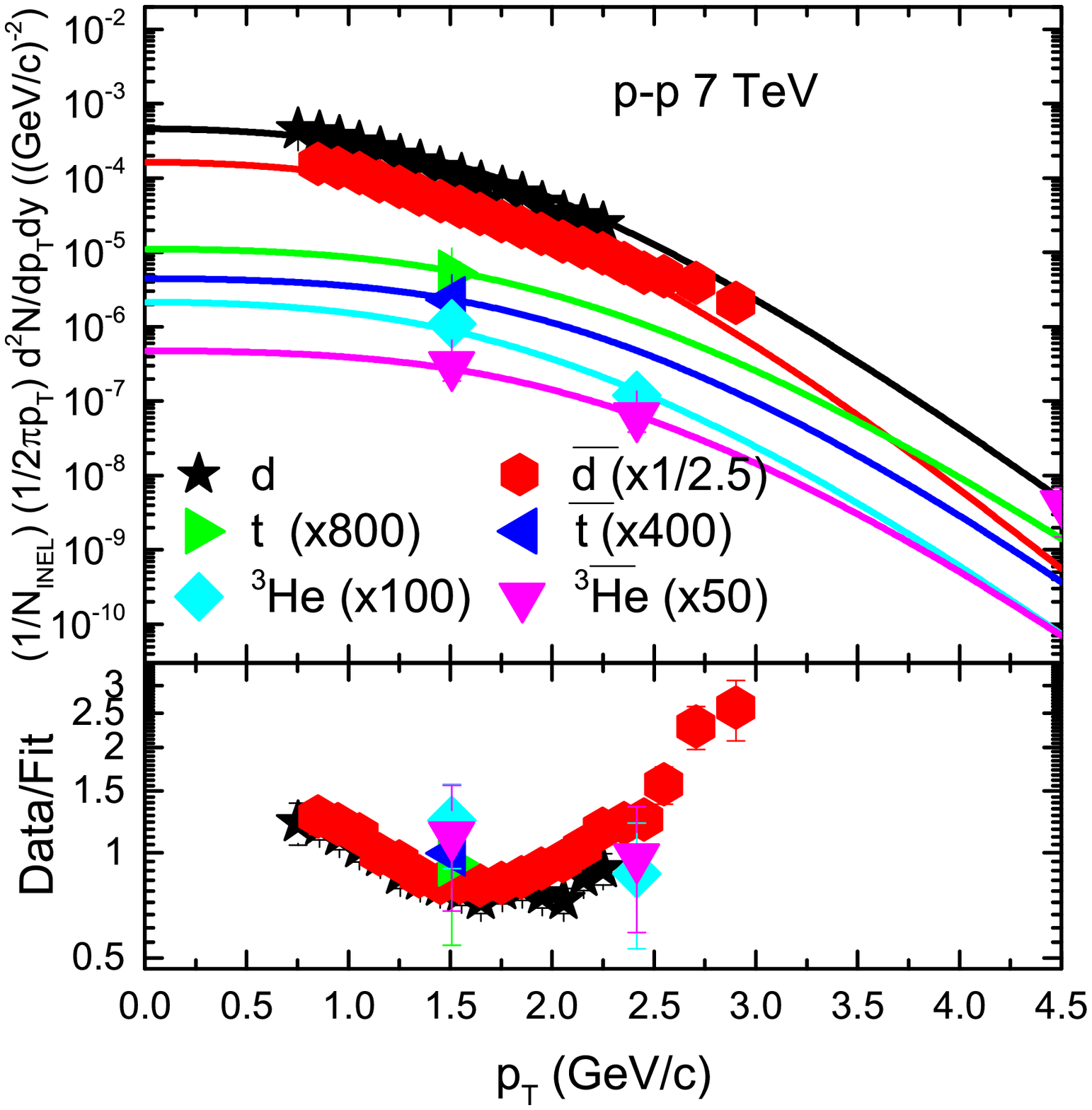}
\end{center}
Fig. 4. is similar to Fig. 2 and 3, but it shows $p_T$ spectra of
$d$, $\bar d$, $t$, $\bar t$, $^3He$ and $\bar {^3He}$ in
$|y|<0.5$ produced in INEL pp collisions at 7 TeV. The symbols
represent the experimental data measured by the ALICE
Collaboration [53], while the curves are our fitted results by
using BGBW statistics, Eqs (1). Each panel is followed by the
ratio of the data/fit.
\end{figure*}
\begin{figure*}[htb!]
\begin{center}
\hskip-0.153cm
\includegraphics[width=15cm]{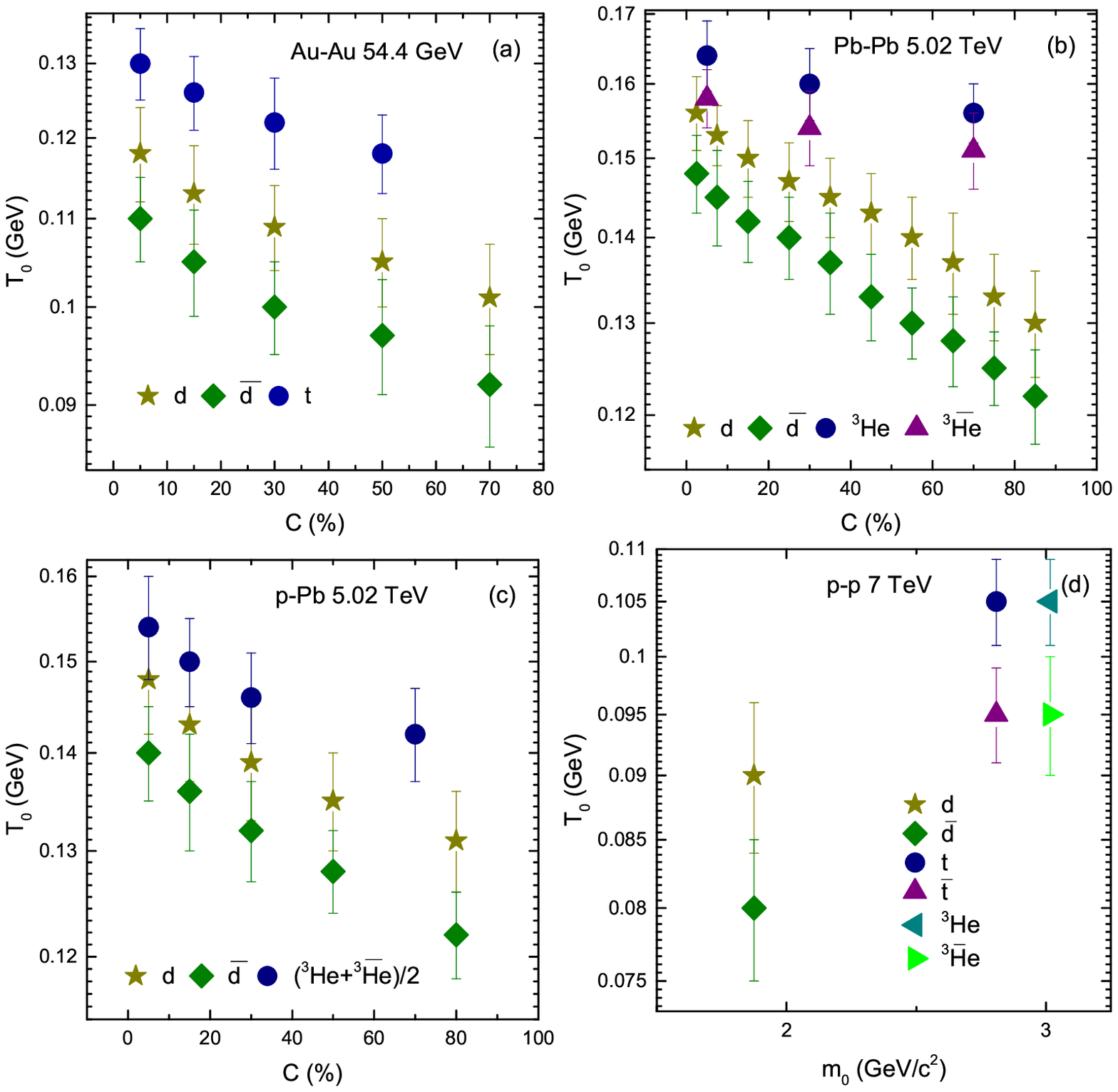}
\end{center}
Fig. 5. Panels (a), (b) and (c): Variation of $T_0$ with
centrality and (d): variation of $T_0$ with $m_0$ for $d$, $\bar
d$, $t$, $\bar t$, $^3He$ and $\bar {^3He}$ or $(^3He+\bar
{^3He})/2$  in Au-Au, Pb-Pb, p-Pb and pp collisions.
\end{figure*}

Fig. 5 includes Panel (a) $d$, $\bar d$, and $t$; Panel (b) $d$,
$\bar d$, $t$ and $\bar t$; Panel (c) $d$, $\bar d$ and
$(^3He+\bar {^3He})/2$; and Panel (d) $d$, $\bar d$, $t$, $\bar
t$, $^3He$ and $\bar {^3He}$. In some panels some particles and
their anti-particles are missing due to the unavailability of
data. In panel (a) one can see that $T_0$ is larger for $t$ than
both of $d$ and $\bar d$ due to its mass. While $d$ and $\bar d$
has the same mass but $d$ freezeout earlier than $\bar d$.
Similarly in panel (b) $d$, $\bar d$, $^3He$ and $\bar {^3He}$,
the mass of $^3He$ and $\bar {^3He}$ is lager than $d$ and $\bar
d$, therefore they freezeout earlier than $d$ and $\bar d$, but
$d$ and $^3He$ freezeout earlier than $\bar d$ and $\bar {^3He}$
respectively, while in panel (c) $(^3He+\bar {^3He})/2$ has larger
$T_0$ than $d$ and $\bar d$ whereas $d$ freezeout earlier than
$\bar d$, and in Panel (d) $^3He$ and $\bar {^3He}$, as well as
$t$ and $\bar t$ freezeout earlier than $d$ and $\bar d$, and the
values for $^3He$ and $t$, and $\bar {^3He}$ and $\bar t$ are
respectively the same. Basically, the formation of light nuclei
occur by the coalescence of nucleons with similar momenta. In the
present work, we believe that the coalescence of nucleons is
larger for $d$, $t$ and $^3He$ compared to their anti-particles
and therefore $T_0$ is larger for light nuclei than for their
anti-nuclei. Furthermore, we observed that $^3He$, and $t$, and
$\bar {^3He}$ and $\bar t$ freezeout at the same time. In our
opinion this is due to the isospin symmetry at high energies which
occurs in nearly identical masses (e.g. triton and helion) where
an up quark is replaced by a down quark. In addition, $T_0$ in
Pb-Pb is larger than in Au-Au and in the later, it is larger than
in pp collisions which shows its dependence on the cross-section
of interacting system. However $T_0$ is larger in p-Pb than in
Au-Au collisions because the center of mass energy for p-Pb is
5.02 TeV which is very larger than the center of mass energy of
Au-Au collision (54.4 GeV), and this may reveal its dependence on
energy.
\begin{figure*}[htb!]
\begin{center}
\hskip-0.153cm
\includegraphics[width=15cm]{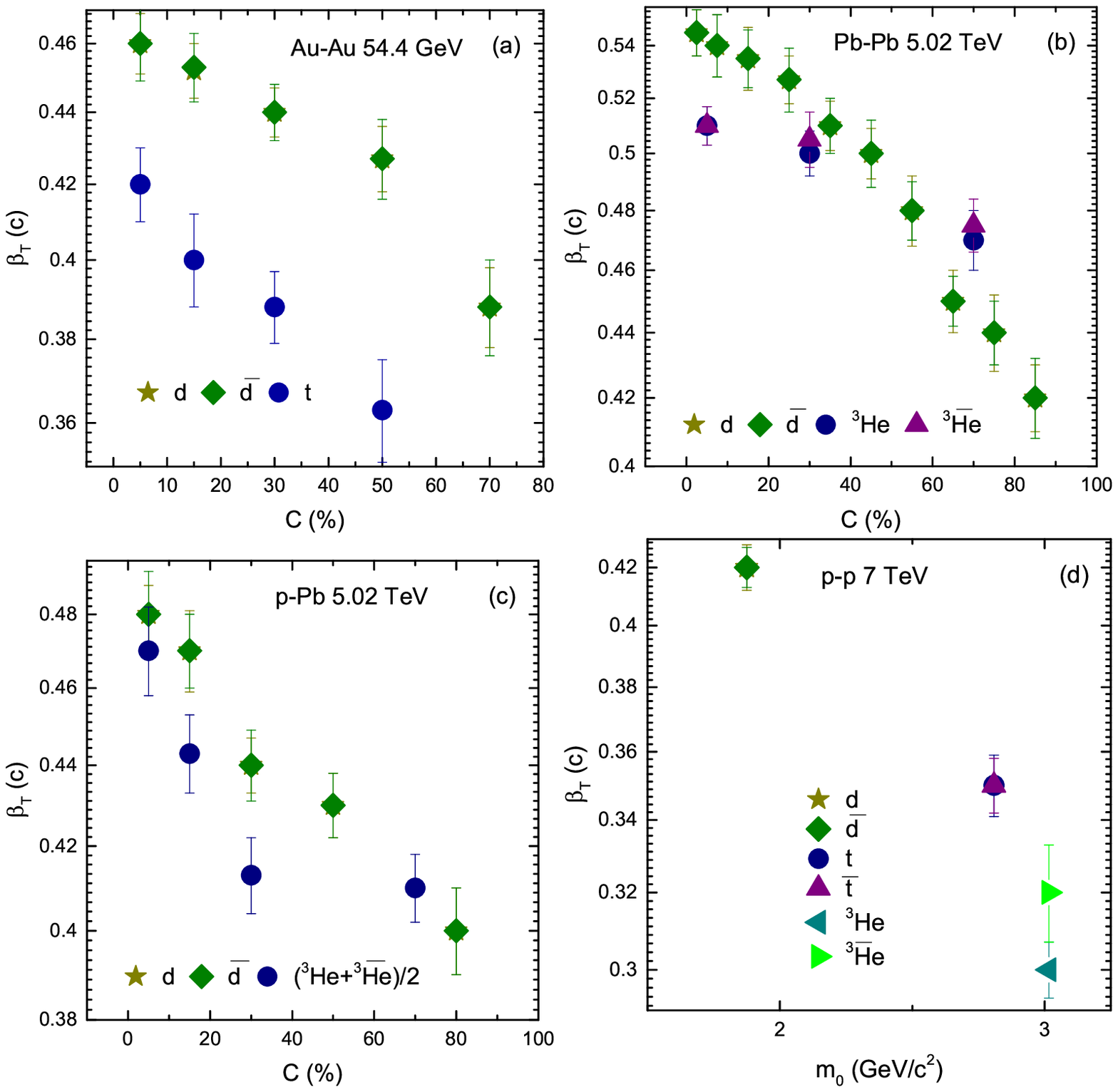}
\end{center}
Fig. 6. Panels (a), (b), (c) Variation of $\beta_T$ with
centrality  and (d) variation of $\beta_T$ with $m_0$ $d$, $\bar
d$, $t$, $\bar t$, $^3He$ and $\bar {^3He}$ or $(^3He+\bar
{^3He})/2$  in Au-Au, Pb-Pb, p-Pb and pp collisions.
\end{figure*}

Fig. 6 is similar to Fig. 5, but shows the dependence of $\beta_T$
on centrality in panels (a), (b) and (c), while panel (d) shows
its dependence on $m_0$. It can be seen that $\beta_T$ decrease
from central to peripheral collisions due to large number of
participants in central collision that experience more violent
squeeze and results in a rapid expansion of the system. While this
expansion becomes steady from central to periphery due to
decreasing the participant nucleons which results in comparatively
weak squeeze. Furthermore, $\beta_T$ is mass dependent. Greater
the mass of the particle is, smaller the value of $\beta_T$.
$\beta_T$ for nuclei and anti-nuclei is the same. Besides,
$\beta_T$ shows dependence on the cross-section of interacting
system. Larger the cross-section of interacting system, larger the
$\beta_T$ is. However $\beta_T$ is slightly larger in p-Pb
collisions than in Au-Au collisions due to the effect of very
large center of mass energy of p-Pb than Au-Au collisions.

\begin{table*}
{\scriptsize Table 1. List of the parameters. (- is used in some
places instead of dof. In fact it is not the fit result. if
dof$<$0, the we put - instead of negative values.) \vspace{-.50cm}
\begin{center}
\begin{tabular}{ccccccccc}\\ \hline\hline
Collisions & Centrality & Particle & $T_0$ (GeV) &$\beta_T$ (c) &$V (fm^3)$      & $q$        & $N_0$ & $\chi^2$/ dof \\ \hline
 Fig. 1        & 0--10\%          & $d$                     &$0.118\pm0.006$  & $0.460\pm0.009$  & $3500\pm183$ & $4.5\times10^{-4}\pm5\times10^{-5}$& 10/11\\
   Au-Au         & 10--20\%         & --                      &$0.113\pm0.006$  & $0.452\pm0.008$  & $3350\pm176$ & $1.7\times10^{-4}\pm5\times10^{-5}$& 4/11\\
   54.4 GeV      & 20--40\%         & --                      &$0.109\pm0.005$  & $0.440\pm0.007$  & $3200\pm150$ & $4\times10^{-5}\pm5\times10^{-6}$  & 11/11\\
                 & 40--60\%         & --                      &$0.105\pm0.005$  & $0.427\pm0.009$  & $3000\pm207$ & $9\times10^{-6}\pm5\times10^{-7}$  & 14/11\\
                 & 60--80\%         & --                      &$0.101\pm0.006$  & $0.388\pm0.010$  & $2800\pm210$ & $1.6\times10^{-6}\pm6\times10^{-7}$ & 17/9\\
\cline{2-8}
  Au-Au           & 0--10\%          & $\bar {d}$              &$0.110\pm0.005$  & $0.460\pm0.011$  & $3200\pm170$  &$5.9\times10^{-5}\pm4\times10^{-6}$ & 12/10\\
  54.4 GeV       & 10--20\%         & --                      &$0.105\pm0.006$  & $0.453\pm0.010$  & $3100\pm180$  & $2.45\times10^{-5}\pm5\times10^{-6}$ & 10/10\\
                 & 20--40\%         & --                      &$0.100\pm0.005$  & $0.440\pm0.008$  & $2900\pm180$  & $7.2\times10^{-6}\pm6\times10^{-7}$  & 22/10\\
                 & 40--60\%         & --                      &$0.097\pm0.006$  & $0.427\pm0.011$  & $2760\pm170$  & $1.6\times10^{-6}\pm4\times10^{-7}$ & 44/10\\
                 & 60--80\%         & --                      &$0.092\pm0.006$  & $0.388\pm0.012$  & $2650\pm140$  & $3.7\times10^{-7}\pm5\times10^{-8}$& 34/8\\
\cline{2-8}
Au-Au            & 0--10\%          & $t$                      &$0.130\pm0.005$  & $0.420\pm0.010$  & $2700\pm190$  & $1.3\times10^{-6}\pm6\times10^{-7}$& 16/3\\
  54.4 GeV       & 10--20\%         & --                       &$0.126\pm0.005$  & $0.400\pm0.012$  & $2500\pm150$  & $5.5\times10^{-7}\pm4\times10^{-8}$& 2/3\\
                 & 20--40\%         & --                      &$0.122\pm0.006$  & $0.388\pm0.009$  & $2360\pm128$  & $1.5\times10^{-7}\pm3\times10^{-8}$& 8/3\\
                 & 40--60\%         & --                      &$0.118\pm0.005$  & $0.363\pm0.012$  & $2200\pm146$  & $3.3\times10^{-8}\pm5\times10^{-9}$& 16/3\\
\hline
Fig. 2           & 0--5\%         & $d$                      &$0.156\pm0.005$  & $0.545\pm0.009$  & $5100\pm172$   & $0.077\pm0.004$        & 1/14\\
  Pb-Pb          & 5--10\%        & --                       &$0.153\pm0.004$  & $0.540\pm0.012$  & $4940\pm160$   & $0.037\pm0.005$        & 3/14\\
  5.02 TeV       & 10--20\%         & --                      &$0.150\pm0.005$  & $0.535\pm0.012$  & $4710\pm160$   & $0.015\pm0.004$        & 4/14\\
                 & 20--30\%         & --                      &$0.147\pm0.005$  & $0.527\pm0.009$  & $4500\pm167$   & $0.006\pm0.0004$       & 5/14\\
                 & 30--40\%         & --                      &$0.145\pm0.005$  & $0.510\pm0.009$  & $4300\pm167$   & $0.0023\pm0.0004$      & 6/14\\
                 & 40--50\%         & --                      &$0.143\pm0.005$  & $0.500\pm0.009$  & $4200\pm167$   & $7.5\times10^{-4}\pm3\times10^{-5}$  & 6/14\\
                 & 50--60\%         & --                      &$0.140\pm0.005$  & $0.480\pm0.012$  & $4100\pm200$   & $2.3\times10^{-4}\pm3\times10^{-5}$ & 21/14\\
                 & 60--70\%         & --                      &$0.137\pm0.006$  & $0.450\pm0.010$  & $3910\pm190$   & $6.3\times10^{-5}\pm4\times10^{-6}$   & 29/11\\
                 & 70--80\%         & --                      &$0.133\pm0.005$  & $0.440\pm0.012$  & $3500\pm160$   & $2.3\times10^{-5}\pm3\times10^{-6}$ & 16/9\\
                 & 80--90\%         & --                      &$0.130\pm0.006$  & $0.420\pm0.010$  & $3300\pm150$   & $3\times10^{-6}\pm5\times10^{-7}$    & 12/6\\
\cline{2-8}
  Pb-Pb             & 0--5\%       & $\bar {d}$               &$0.148\pm0.005$  & $0.545\pm0.009$  & $4700\pm181$   & $0.086\pm0.005$        & 3/14\\
  5.02 TeV       & 5--10\%         & --                       &$0.145\pm0.006$  & $0.540\pm0.012$  & $4500\pm190$   & $0.04\pm0.006$         & 5/14\\
                 & 10--20\%        & --                       &$0.142\pm0.005$  & $0.535\pm0.011$  & $4300\pm172$   & $0.0166\pm0.004$       & 5/14\\
                 & 20--30\%        & --                       &$0.140\pm0.005$  & $0.527\pm0.012$  & $4100\pm150$   & $0.007\pm0.0004$       & 6/14\\
                 & 30--40\%        & --                       &$0.137\pm0.006$  & $0.510\pm0.010$  & $3900\pm170$   & $0.0026\pm0.0004$       & 9/14\\
                 & 40--50\%        & --                       &$0.133\pm0.005$  & $0.500\pm0.012$  & $3700\pm180$   & $8.5\times10^{-4}\pm5\times10^{-5}$ & 12/14\\
                 & 50--60\%        & --                       &$0.130\pm0.004$  & $0.480\pm0.010$  & $3500\pm156$   & $2.6\times10^{-4}\pm4\times10^{-5}$ & 26/14\\
                 & 60--70\%        & --                       &$0.128\pm0.005$  & $0.450\pm0.008$  & $3300\pm160$   & $7.2\times10^{-5}\pm6\times10^{-6}$ & 22/11\\
                 & 70--80\%        & --                       &$0.125\pm0.004$  & $0.440\pm0.010$  & $3100\pm180$   & $2.6\times10^{-5}\pm5\times10^{-6}$ & 16/9\\
                 & 80--90\%        & --                       &$0.122\pm0.005$  & $0.420\pm0.012$  & $2900\pm140$   & $3.2\times10^{-6}\pm4\times10^{-7}$   & 15/6\\
\hline
Pb-Pb            & 0--10\%         & $^3He$                   &$0.164\pm0.005$  & $0.510\pm0.007$  & $4400\pm160$   &$1.5\times10^{-7}\pm4\times10^{-8}$ & 3/1\\
  5.02 TeV       & 10--40\%         & --                      &$0.160\pm0.005$  & $0.500\pm0.008$  & $4200\pm140$   &$9\times10^{-8}\pm4\times10^{-9}$ & 2/1\\
                 & 40--100\%         & --                     &$0.156\pm0.004$  & $0.470\pm0.010$  & $4000\pm180$   &$4.7\times10^{-8}\pm4\times10^{-9}$ & 10/1\\
\hline
Pb-Pb            & 0--10\%          & $\bar {^3He}$           &$0.158\pm0.004$  & $0.510\pm0.007$  & $4000\pm140$   &$1.4\times10^{-7}\pm4\times10^{-8}$ & 4/1\\
  5.02 TeV       & 10--40\%         & --                      &$0.154\pm0.005$  & $0.505\pm0.010$  & $3800\pm166$   &$9.5\times10^{-8}\pm4\times10^{-8}$ & 3/1\\
                 & 30--100\%         & --                     &$0.151\pm0.005$  & $0.475\pm0.009$  & $3600\pm155$   &$5.3\times10^{-8}\pm4\times10^{-9}$ & 4/1\\
\hline
Fig. 3            & 0--10\%          & $d$                     &$0.148\pm0.006$  & $0.480\pm0.008$  & $4720\pm170$   &$1.25\times10^{-4}\pm4\times10^{-5}$& 12/7\\
  p-Pb           & 10--20\%         & --                      &$0.143\pm0.006$  & $0.470\pm0.011$  & $4550\pm160$   &$4.5\times10^{-5}\pm5\times10^{-6}$ & 18/7\\
  5.02 TeV       & 20--40\%         & --                      &$0.139\pm0.006$  & $0.440\pm0.007$  & $4400\pm186$   &$1.7\times10^{-5}\pm6\times10^{-6}$ & 13/4\\
                 & 40--60\%         & --                      &$0.135\pm0.005$  & $0.430\pm0.008$  & $4200\pm179$   &$5\times10^{-6}\pm5\times10^{-7}$   & 21/4\\
                 & 60--100\%         & --                     &$0.131\pm0.005$  & $0.400\pm0.010$  & 4130$\pm180$   &$8\times10^{-7}\pm6\times10^{-8}$   & 35/4\\
\hline
p-Pb             & 0--10\%          & $\bar {d}$              &$0.140\pm0.005$  & $0.480\pm0.012$  & $4400\pm184$   &$1.31\times10^{-4}\pm6\times10^{-5}$ &8/7\\
  5.02 TeV       & 10--20\%         & --                      &$0.136\pm0.006$  & $0.470\pm0.010$  & $4200\pm140$   &$4.4\times10^{-5}\pm6\times10^{-6}$ & 18/5\\
                 & 20--40\%         & --                      &$0.132\pm0.005$  & $0.440\pm0.009$  & $4000\pm160$   &$1.75\times10^{-5}\pm7\times10^{-6}$ & 8/5\\
                 & 40--60\%         & --                      &$0.128\pm0.004$  & $0.430\pm0.008$  & $3800\pm154$   &$5.5\times10^{-6}\pm7\times10^{-7}$ & 21/4\\
                 & 60--100\%         & --                     &$0.122\pm0.004$  & $0.400\pm0.010$  & 3600$\pm168$   &$1\times10^{-6}\pm4\times10^{-7}$   & 16/4\\
\hline
p-Pb             & 0--10\%          & $(^3He+\bar {^3He})/2$  &$0.154\pm0.006$  & $0.470\pm0.008$  & $4000\pm191$   &$5\times10^{-8}\pm4\times10^{-9}$ & 4/-\\
5.02 TeV       & 10--20\%         & --                      &$0.150\pm0.005$  & $0.443\pm0.011$  & $3800\pm166$   &$1.4\times10^{-8}\pm5\times10^{-9}$ & 3/-\\
                 & 20--40\%         & --                      &$0.146\pm0.005$  & $0.413\pm0.009$  & $3600\pm165$   &$5\times10^{-9}\pm3\times10^{-10}$  & 26/-\\
              & 40--100\%         & --                     &$0.142\pm0.005$  & $0.410\pm0.011$  & 3400$\pm150$   &$7\times10^{-10}\pm3\times10^{-11}$& 1/-\\
\hline
Fig.4           & --               & $d$                       &$0.090\pm0.006$  & $0.420\pm0.008$  & $3000\pm158$   &$2\times10^{-6}\pm4\times10^{-7}$ & 141/17\\
pp              & --               & $\bar {d}$              &$0.080\pm0.005$  & $0.420\pm0.007$  & $2600\pm145$   &$2\times10^{-6}\pm4\times10^{-7}$ & 81/16\\
  7 TeV         & --               & $t$                     &$0.105\pm0.004$  & $0.350\pm0.009$  & $2000\pm155$   &$1.3\times10^{-10}\pm3\times10^{-11}$ & 0.02/-\\
                & --               & $\bar {t}$                &$0.095\pm0.004$  & $0.350\pm0.008$  & $2000\pm145$   &$1.6\times10^{-10}\pm4\times10^{-11}$ & 0.02/-\\
                & --               & $^3He$                  &$0.105\pm0.004$  & $0.300\pm0.007$  & $2000\pm155$   &$1.3\times10^{-10}\pm3\times10^{-11}$&0.6/-\\
                & --               & $\bar {^3He}$           &$0.095\pm0.005$ & $0.350\pm0.007$    & $2000\pm145$   &$6\times10^{-11}\pm4\times10^{-12}$  & 38/-\\
\hline
\end{tabular}%
\end{center}}
\end{table*}

\begin{figure*}[htb!]
\begin{center}
\hskip-0.153cm
\includegraphics[width=15cm]{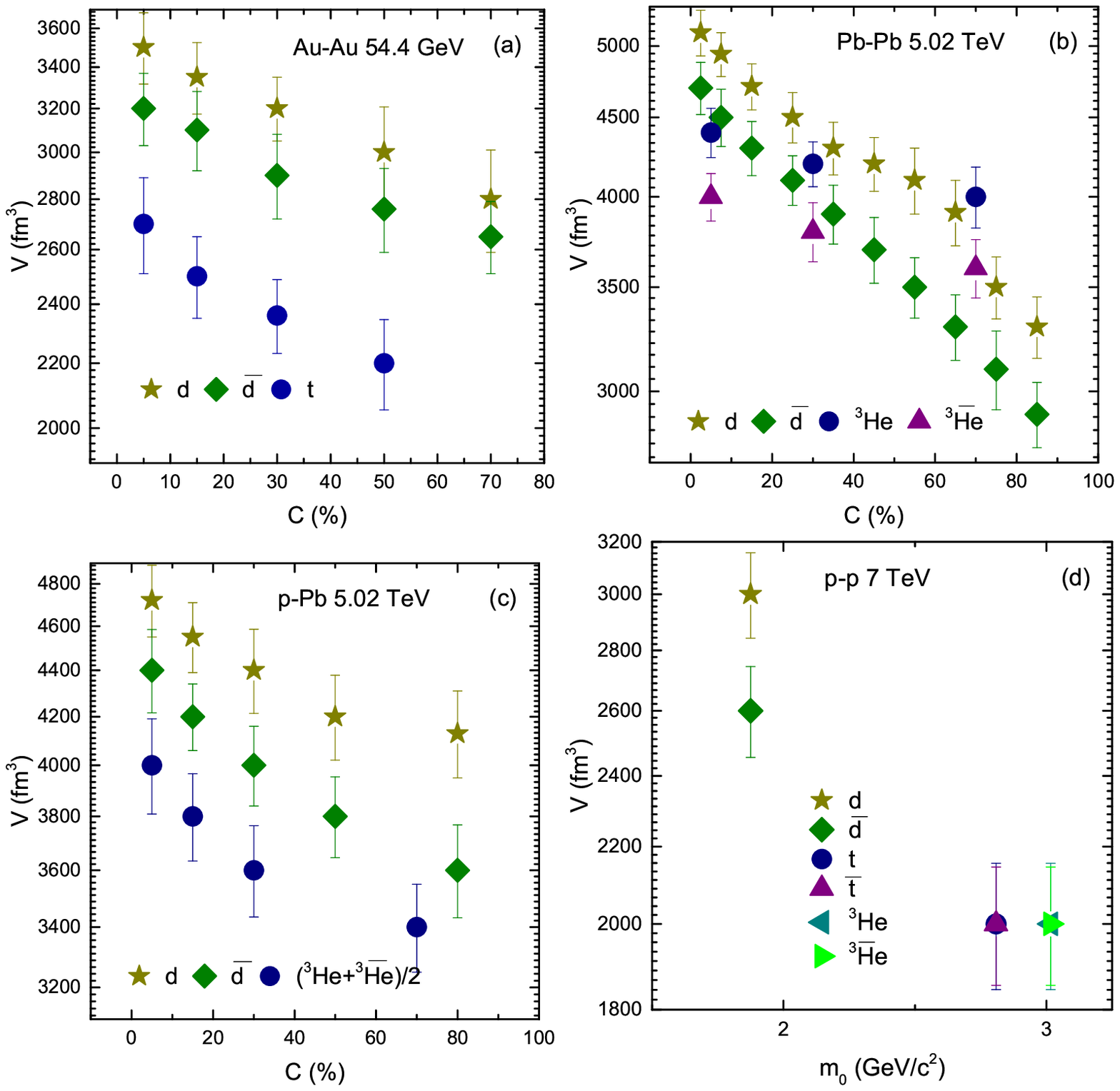}
\end{center}
Fig.7 Panels (a), (b), (c) Variation of $V$ with centrality and
(d) variation of $\beta_T$ with $m_0$ for $d$, $\bar d$, $t$,
$\bar t$, $^3He$ and $\bar {^3He}$ or $(^3He+\bar {^3He})/2$  in
Au-Au, Pb-Pb, p-Pb and pp collisions.
\end{figure*}

Fig. 7 is similar to Fig. 6, but shows the dependence of $V$ on
centrality (mass). It can be seen that $V$ decreases from central
to peripheral collisions in panel (a), (b) and (c), as the number
of participant nucleons decreases from central collisions to
periphery depending on the interaction volume. The system with
more participants reaches to equilibrium quickly due to large
number of secondary collisions by the re-scattering of partons,
which decreases towards periphery and the system goes away from
equilibrium state. In addtion, $V$ for deuteron and anti-deuteron
is larger than that of triton and anti-triton as well as from
helion and anti-helion. The parameter $V$ for nuclei is larger
than their anti-nuclei due to larger coalescence of nucleons for
the nuclei than for their anti-nuclei. In case of triton and
anti-triton and helion and anti-helion, V of triton and helion,
and anti-triton and anti-helion are respectively the same.
Besides, $V$ is larger in Pb-Pb collisions that the rest, and in
Au-Au collisions as well as p-Pb collisions it is larger than in
pp collisions which shows the dependence of $V$ on the
cross-section of interacting system. However $V$ is larger in p-Pb
than in Au-Au collisions. We think that this is due the effect of
very higher center energy of p-Pb collisions compared to Au-Au
collisions, because higher energy corresponds to longer evolution
time which may lead to larger partonic system.

\begin{figure*}[htb!]
\begin{center}
\hskip-0.153cm
\includegraphics[width=15cm]{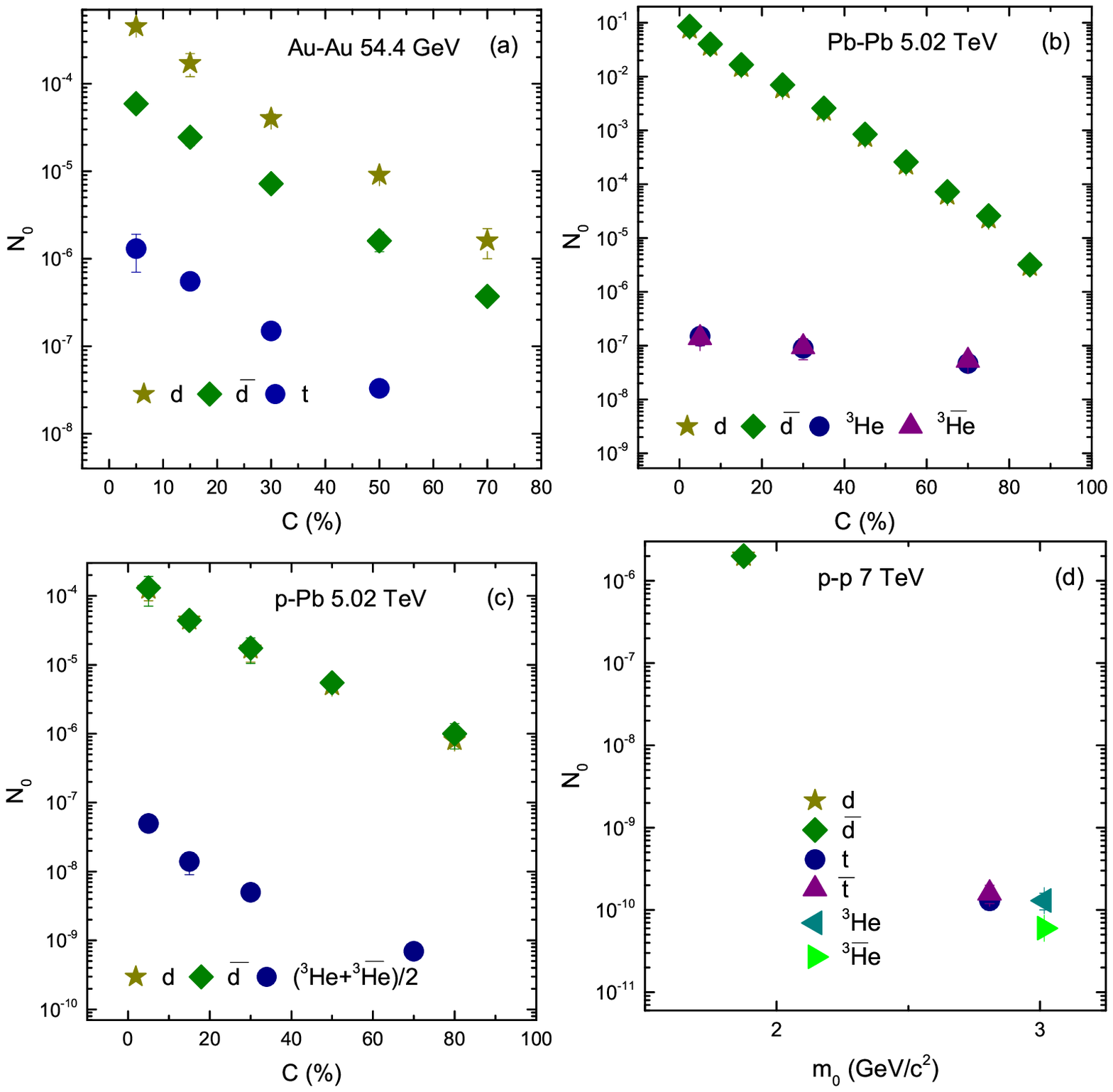}
\end{center}
Fig.8 Panels (a), (b), (c) Dependence of $N_0$ on centrality and
(d) Dependence of $N_0$ on $m_0$, for $d$, $\bar d$, $t$, $\bar
t$, $^3He$ and $\bar {^3He}$ or $(^3He+\bar {^3He})/2$  in Au-Au,
Pb-Pb, p-Pb and pp collisions.
\end{figure*}

Figs. 8 (a), (b) and (c) show the dependence of $N_0$ on
centrality. One can see that $N_0$ decrease with decreasing
centrality. Furthermore, the parameter $N_0$ depends on mass of
the particle. In panels (a)-(d) the parameter $N_0$ for deuteron
and anti-deuteron are larger than triton, and in (a) it is larger
for deuteron than anti-deuteron due to large coalescence of
deuteron, while in panels (b)-(d) the parameter $N_0$ for nuclei
and their anti-particles are the almost same. In general, the
parameter $N_0$ is the same for nuclei and their anti-nuclei. The
parameter $N_0$ for triton and helion, and anti-triton and
anti-helion is the same due to the isopin symmetry. In deed $N_0$
is only a normalization factor and the data are not cross-section,
but they are proportional to the volumes of sources of producing
various particles. Therefore it is significant to study $N_0$
dependence.

\begin{figure*}[htb!]
\begin{center}
\hskip-0.153cm
\includegraphics[width=15cm]{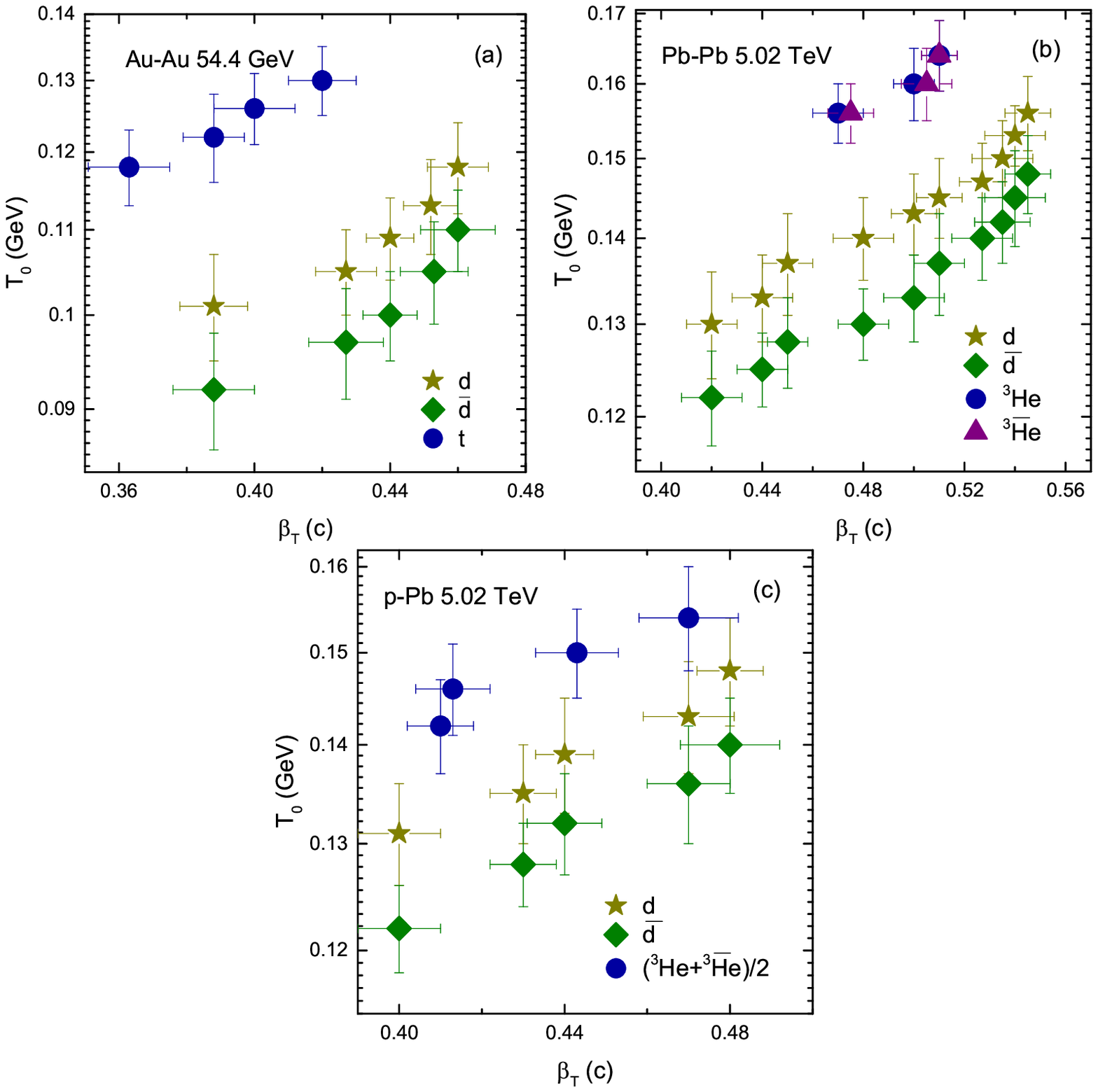}
\end{center}
Fig.9 Variation of $T_0$ with $\beta_T$.
\end{figure*}

Fig. 9 shows the variation of $T_0$ with $\beta_T$. It is observed
that central collisions correspond to larger $T_0$ and $\beta_T$.
The correlation between $T_0$ and $\beta_T$ is positive. This is
in agreement with [16] and in disagreement with  [18]. In panel
(a) the correlation of $T_0$ and $\beta_T$ is larger for triton
and that of deuteron is larger than anti-deuteron. In panel (b)
the correlation of $T_0$ and $\beta_T$ is larger for helions than
deuterons and that of deuteron is larger for than anti-deuteron.
Similarly in panel (c) the helions has larger correlation between
$T_0$ and $\beta_T$ than deuteron and anti-deuteron, and for
deuteron it is larger than anti-deuteron. In general, the massive
particles has larger correlation between of $T_0$ and $\beta_T$,
and the particles has larger correlation than their anti-particles
due to less coalescence of anti-particles.

\begin{figure*}[htb!]
\begin{center}
\hskip-0.153cm
\includegraphics[width=15cm]{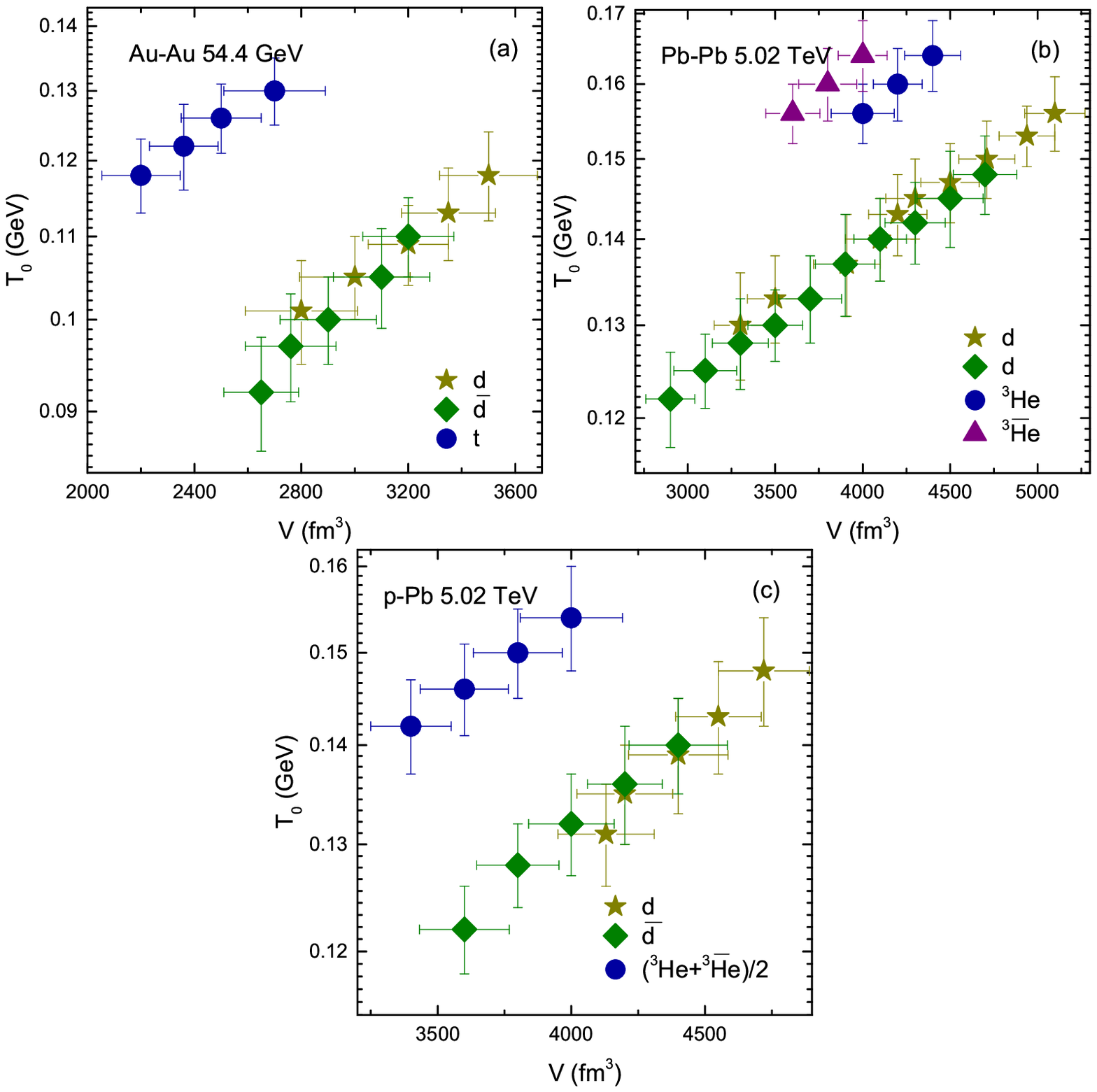}
\end{center}
Fig.10 Variation of $T_0$ with $V$.
\end{figure*}

Fig. 10 shows the variation of $T_0$ with $V$. It is observed that
central collisions correspond to larger $T_0$ and $V$. The
correlation between $T_0$ and $V$ is positive. In panel (a) the
correlation of $T_0$ and $V$ is larger for triton, while the
correlation of $T_0$ and $V$ for deuterons is slightly larger than
anti-deuteron. In panel (b) the correlation of $T_0$ and $V$ is
larger for helions than deuterons, and that of helions and
deuterons is slightly larger for than their anti-particles.
Similarly in panel (c) the helions has larger correlation between
$T_0$ and $V$ than deuteron and anti-deuteron and deuteron has
larger correlation of $T_0$ and $V$ than anti-deuteron. In short
the massive particles has larger correlation between of $T_0$ and
$V$, and the particles has larger correlation than their
anti-particles due to less coalescence of anti-particles.
\\
 {\section{Conclusions}}
 The main observations and conclusions are summarized here.

(a) The transverse momentum spectra of light nuclei and their
anti-nuclei produced in Au-Au, Pb-Pb and p-Pb and produced in
inelastic pp collisions in different centrality intervals are
analyzed by the BGBW model. The model results show an agreement
with the experimental data in the special $p_T$ range measured by
the STAR and ALICE Collaborations.

(b) Kinetic freezeout temperature is larger for triton and helions
and their anti-particles than deuteron and anti-deuteron due to
their mass. While helions and tritons have the same value of
Kinetic freezeout temperature due to isospin symmetry, and
dueteron, triton and helion freezeout earlier than their
anti-particles respectively due to large coalescence of nucleons
for the light nuclei than anti-nuclei.

(c) Kinetic freezeout temperature decrease from central to
peripheral collisions due to the decrease of participant nucleons
in the peripheral collisions which lead to the decrease in the
degree of excitation of the system.

d) Transverse flow velocity increases from peripheral to central
collisions due the reason that the collisions become more violent
in central collisions which also expands the system rapidly.

e) The kinetic freezeout volume decreases from central to
peripheral collisions and the system reaches quickly to
equilibrium state due to large number of secondary collisions by
the re-scattering of partons in central collisions that decreases
towards periphery. In addition, The normalization constant is
larger in central collisions than in peripheral collisions.

   )
\\
\\

{\bf Data availability}

The data used to support the findings of this study are included
within the article and are cited at relevant places within the
text as references.
\\
\\
{\bf Compliance with Ethical Standards}

The authors declare that they are in compliance with ethical
standards regarding the content of this paper.
\\
\\
{\bf Acknowledgements}

The authors would like to thank support from the National Natural Science
Foundation of China (Grant Nos. 11875052, 11575190, and 11135011).
\\
\\

\end{multicols}
\end{document}